\documentstyle[aps,manuscript]{revtex}
\begin{document}
\tightenlines
\title{Inflation from a massive scalar field and scalar perturbations
of the metric}
\author{F. Lara, F. Astorga, M. Bellini}
\address{
Instituto de F\'{\i}sica y Matem\'aticas,
Universidad Michoacana de San Nicol\'as de Hidalgo,\\
AP:2-82, (58041) Morelia, Michoac\'an, M\'exico.}
\maketitle
\begin{abstract}
In the framework of chaotic inflation we study the case of a
massive scalar field using  a semiclassical approach.
We derive the energy density fluctuations in the infrared sector
generated by matter field and gauge-invariant metric fluctuations
by means of a different method. We find that the super Hubble 
density fluctuations increase during inflation for a massive 
scalar field, in agreement with the result obtained in 
a previous work in which a power-law expanding universe was
considered.
\end{abstract}
%%%%%%%%%%%%%%%%%%%%%%%%%%%%%%%%%%%%%%%%%%%%%%%%%%%%%%%%%%%%%%%%%%%%%%%%%%
\section{Introduction}

Inflationary cosmology was developed by Guth in the 80's\cite{Guth}
in order to solve some of the shortcomings of the big bang theory, and in
particular to explain the extraordinary homogeneity of the observable 
universe. However, the universe after inflation in this scenario
becomes very inhomogeneous. Following a detailed investigation
of this problem, A. Guth and E. Weinberg concluded that the old
inflation model could not be improved\cite{GW}. 

These problems were sorted out by A. Linde in 1983
with the introduction of chaotic inflation\cite{lin}.
In this scenario inflation can occur in theories with
potentials such as $V(\varphi) \sim \varphi^n$. It may
begin in the absence of thermal equilibrium in the early
universe, and it may start even at the Planckian density, in
which case the problem of initial conditions for inflation
can be easily solved\cite{libro}. According to the
simplest versions of chaotic inflationary theory the 
universe is not a single expanding ball of fire produced in the big
bang, but rather a huge eternally growing fractal. It consists
of many inflating balls that produce new balls, which produce
more new balls, ad infinitum.

Within this context, the Cosmic Microwave Background Radiation  
plays a fundamental role. Data coming from COBE and the other  
CMB radiation detectors have started shifting inflation from a mainly 
theoretical subject to a more constrained field. It is expected 
that new observations from the MAP, to be launched this summer, 
and Planck missions, planned for 2007, are crucial for the understanding
of an inflationary stage in the universe. With the advent of new 
CMBR physics, it is thought that the measurements of the power 
spectrum of temperature fluctuations at much smaller angular separations, 
of the order of fractions of a degree, might be sensitive to the standard 
model parameters with unprecedented precision. This will bring a 
remarkable change in the field. 

The dynamics of a scalar field $\varphi $
minimally coupled to a classical gravitational
one is described by the Lagrangian:
\begin{equation}
{\cal L}(\varphi,\varphi_{,\mu})=-\sqrt{-g}\left[ \frac{R}{16\pi}+
\frac{1}{2}g^{\mu\nu}\varphi_{,\mu}\varphi_{,\nu} + V(\varphi)\right] \ .
\end{equation}
If the spacetime has a Friedman-Robertson-Walker (FRW) metric, 
$ds^2=-dt^2+a^2(t) dr^2$, which describes a globally 
isotropic and homogeneous universe,
the equations of motion for the field
operator $\varphi$ and the Hubble parameter $H\equiv\frac{\dot a}{a}$ are:
\begin{eqnarray}
&&\ddot\varphi-\frac{1}{a^2}\nabla^2\varphi+3H\dot\varphi+V'(\varphi)=0 \ ,\\
&&H^2=\frac{4\pi}{3M_p^2}\left < \dot\varphi^2
+\frac{1}{a^2}(\vec\nabla\varphi)^2 +2V(\varphi) \right > \label{3} ,
\end{eqnarray}
where the overdot denotes the time derivative and $V'(\varphi)\equiv
\frac{d V}{d\varphi}$. The expectation value is assumed to be a constant
function of the spatial variables for consistency with the FRW metric.  
When inflation ends the field starts oscillating rapidly and its
potential energy is converted into thermal energy.  This is the general
scheme of the inflationary scenario without considering the quantum effects.

In this work we consider a
semiclassical expansion of the theory. To obtain this we decompose
the inflaton operator in a classical field $\phi_c$ plus the 
quantum fluctuations
$\phi$, $\varphi=\phi_c+\phi$. 
The spatially homogeneous field $\phi_c(t)$ is defined as the solution
to the classical equation of motion [here, $H(\varphi) \equiv H_c(\phi_c)$]:
\begin{equation}    \label{4}
\ddot \phi_c+3H_c \dot\phi_c+V'(\phi_c)=0,
\end{equation}
where the prime denotes the derivative with respect to $\phi_c$.
The evolution of the quantum
operator $\phi$ is given by:
\begin{equation}    \label{5}
\ddot \phi-\frac{1}{a^2}\nabla^2 \phi +3H_c(\phi_c)\dot\phi+V''(\phi_c) \phi=0,
\end{equation}
where $H_c(\phi_c)={\dot a \over a}$ is the Hubble
parameter and the classical Friedmann equation
is
\begin{equation}   \label{7}
H_c^2=\frac{4\pi}{3M_p^2}\left[\dot \phi_c^2+ 2V(\phi_c)\right].
\end{equation}
From eqs.(\ref{4}) and (\ref{7}), we obtain
the classical dynamics
of the Hubble parameter and the inflaton field 
given by the following relations:
\begin{eqnarray}
\dot \phi_c &=& -\frac{M_p^2}{4\pi} H_c'\ , \label{9}\\
\dot H_c & = & H_c'\dot\phi_c =-\frac{M_p^2}{4\pi} (H_c')^2 \ .\label{10}
\end{eqnarray}
Hence, from eq. (\ref{3}) we derive the expression for the
scalar potential $V(\phi_c)$ 
\begin{equation}       \label{11}
V(\phi_c)=\frac{3M_p^2}{8\pi}\left(H_c^2-
\frac{M_p^2}{12\pi}(H_c')^2\right) \label{p2} \ .
\end{equation}
Equations (\ref{9}) and (\ref{10}) define the classical evolution of
space-time, and determine the relation between the classical 
potential and the inflationary regimes. 
On the other hand eq.(\ref{5}) defines the quantum
dynamics of the field $\phi$.

If during inflation the slow-roll condition is impossed, the following
conditions are required\cite{Liddle}
\begin{eqnarray}
\gamma &=& \frac{2}{\kappa^2} \left(\frac{H'_c}{H_c}\right)^2, \label{a}\\
\eta & = &   \frac{2}{\kappa^2} \frac{H''_c}{H_c} \label{b},
\end{eqnarray}
where $\kappa^2 = {8\pi \over M^2_p}$ and $M_p \simeq 1.2 10^{19} \  GeV$ 
is the Planckian mass. Furthermore, to solve the flatness problem
$60$ or more e-folds are required.The number of e-folds
during inflation is
\begin{equation}
N = \frac{a(t_f)}{a(t_i)}=\int^{t_f}_{t_i} H_c(t) \  dt,
\end{equation}
where $a$ is the scale factor of the universe and $t_i$ and $t_f$ are
respectively the times when inflation starts and ends.

In this work we consider an inflaton field with mass 
$m=10^{-5} \  M_p$.

\section{Gauge-invariant metric fluctuations}

The issue of gauge-invariance becomes critical when we attempt to
analyze how the scalar metric perturbations produced in the very
early universe influence a globally flat FRW background metric. For a
diagonal stress $T_{ij}$ the perturbed FRW metric is described
by\cite{Bardeen}
\begin{equation}
ds^2=\left(1+2\psi\right) dt^2 - a^2(t) \left(1-2\psi\right) dx^2,
\end{equation}
where $\psi$ are the perturbations of the metric.

After linearizing the Einstein equations in terms of $\phi$ and $\psi$,
one obtains
\begin{eqnarray}
\ddot\psi &+& \left(\frac{\dot a}{a}
- 2 \frac{\ddot\phi_c}{\dot\phi_c} \right)
\dot \psi - \frac{1}{a^2} \nabla^2 \psi +2\left[
\frac{\ddot a}{a} - \left(\frac{\dot a}{a}\right)^2 - \frac{\dot a}{a}
\frac{\ddot\phi_c}{\dot\phi_c}\right] \psi =0, \label{1}\\
\frac{1}{a}& \frac{d}{dt}& \left( a \psi \right)_{,\beta} = 
\frac{4\pi}{M^2_p} \left(\dot\phi_c \phi\right)_{,\beta} , \\
\ddot\phi& +& 3 \frac{\dot a}{a} \dot\phi -
\frac{1}{a^2} \nabla^2 \phi + V''(\phi_c) \phi 
+ 2 V'(\phi_c) \psi- 4 \dot\phi_c \dot\psi =0, \label{A}
\end{eqnarray}
where the dynamics of $\phi_c$ is given by the eqs. (\ref{4}) and
(\ref{9}). Eq. (\ref{1}) can be simplified by introducing
the field $h = e^{1/2\int\left[\dot a/a - 2\ddot\phi_c/\phi_c\right]dt}
\psi$
\begin{eqnarray}
\ddot h &-& \frac{1}{a^2} \nabla^2 h -  \left[ \frac{1}{4}
\left(\frac{\dot a}{a} - 2 \frac{\ddot \phi_c}{\dot\phi_c} \right)^2
+\frac{1}{2} \left( \frac{\ddot{a} a - \dot{a}^2}{a^2}
- \frac{2 \frac{d}{dt}\left(\ddot\phi_c \dot\phi_c\right)
- 4 \dot\phi^2_c}{\dot\phi^2_c} \right)\right.\nonumber\\
& -& 2\left( \frac{\ddot a}{a} - \left( \frac{\dot a}{a}\right)^2
- \left.
\frac{\dot a}{a} \frac{\ddot\phi_c}{\dot\phi_c} \right) \right] h
= 0\label{h}.
\end{eqnarray}
This field can be expanded in terms of the modes
$h_k=e^{i\vec k.\vec x} \tilde\xi_k(t)$
\begin{equation}
h(\vec x,t) = \frac{1}{(2\pi)^{3/2}} {\Large \int} d^3k \left[
\alpha_k h_k + \alpha^{\dagger}_k h^*_k\right],
\end{equation}
where $\alpha_k$ and $\alpha^{\dagger}_k$ are the annihilation
and creation operators with commutation relations
\begin{eqnarray}
\left[\alpha_k,\alpha^{\dagger}_{k'} \right]&=& \delta^{(3)}(k-k'), \label{c1}\\
\left[\alpha_k,\alpha_{k'} \right]&=&\left[
\alpha^{\dagger}_k,\alpha^{\dagger}_{k'} \right]=0.\label{c2}
\end{eqnarray}
The equation for the modes $\tilde\xi_k$ is
\begin{equation}\label{xxi}
\ddot{\tilde\xi}_k + \tilde\omega^2_k(t) \  \tilde\xi_k =0,
\end{equation}
where $\tilde\omega^2_k = k^2/a^2 - \tilde k^2_0/a^2$ is the squared
time dependent frequency and $\tilde k_0$ separates the infrared
and ultraviolet sectors, and is given by 
\begin{equation}
\frac{\tilde k^2_0}{a^2} = \frac{1}{4} \left(\frac{\dot a}{a}
- 2 \frac{\ddot\phi_c}{\dot\phi_c}\right)^2 +\frac{1}{2}\left(
\frac{\ddot a a - \dot a^2}{a^2} 
- \frac{2\frac{d}{dt}\left(\ddot\phi_c \dot\phi_c\right)
- 4 \ddot\phi^2_c}{\dot\phi^2_c}\right) 
-2 \left[\frac{\ddot a}{a} - \left(\frac{\dot a}{a}\right)^2
-\frac{\dot a}{a} \frac{\ddot\phi_c}{\dot\phi_c} \right].
\end{equation}
Note that for super Hubble scales ($k^2 \ll \tilde k^2_0$), eq.
(\ref{xxi})  can be approximated by
\begin{equation}\label{C}
\ddot{\tilde\xi}_0 - \frac{\tilde k^2_0}{a^2} \  \tilde\xi_0 =0,
\end{equation}
where $\tilde\xi_0$ denotes the zero-mode $\tilde\xi_{k=0}(t)$.

Finally, the fluctuations for the energy density are\cite{li}
\begin{equation}\label{22}
\frac{\delta\rho}{\rho} = -2 \psi.
\end{equation}

\section{The massive scalar field}

We consider a massive scalar field with potential 
\begin{equation}
V(\phi_c) = \frac{m^2}{2} \phi^2_c.
\end{equation}
If the slow rolling conditions (\ref{a}) and (\ref{b}) are fulfilled, the
second term inside the brackets of eq. (\ref{11}) can be neglected, so that
the Hubble parameter is given by
\begin{equation}\label{H}
H_c(\phi_c) = 2\sqrt{\frac{\pi}{3}} \frac{m}{M_p} \  \phi_c.
\end{equation}
From eq. (\ref{9}) one obtains the time evolution for the
scalar field
\begin{equation}\label{c}
\phi_c(t) = \phi_i - \frac{m M_p}{2 \sqrt{3\pi}} t,
\end{equation}
where $\phi_i \equiv \phi_c(t_i)$ is the value of the homogeneous field
when inflation starts. The scale factor, written as a function 
of $\phi_c$, is expressed as 
\begin{equation}
a(\phi_c) = a_i \  e^{-\frac{2\pi}{M^2_p} \phi^2_c},
\end{equation}
and increases with time due to the fact that $\phi_c$ decreases with time.
Here, $a_i \equiv a(t_i)$.
From eqs. (\ref{H}), (\ref{a}) and (\ref{b}) one obtains
the slow roll parameters
\begin{eqnarray}
\gamma & = & \frac{M^2_p}{4\pi} \  \phi^{-2}_c,\\
\eta & = & 0,
\end{eqnarray}
which complies with the required conditions for slow-rolling 
for $\phi_c \gg {M_p \over 2\sqrt{\pi}}$.

\subsection{Super Hubble matter field fluctuations}

The equation of motion (\ref{5})
for the matter field fluctuations $\phi$ can be simplified by means
of the change of variable $\chi = a^{3/2} \phi$. 
With this transformation the following equation
for the redefined field $\chi$ becomes 
\begin{equation}
\ddot\chi + \left[\frac{k^2}{a^2} - \frac{k^2_0}{a^2}\right]\chi=0,
\end{equation}
where $k_0$ is the wavenumber that separates the infrared ($k^2 \ll k_0$)
and ultraviolet ($ k^2 \gg k^2_0$) sectors. The squared time dependent
parameter of mass $\mu^2(t) = {k^2_0 \over a^2}$ is given by
\begin{equation}
\mu^2(t) = \frac{9}{4} H^2_c + \frac{3}{2} \dot H_c - V''_c.
\end{equation}

The redefined matter field fluctuations $\chi(\vec x,t)$ for 
the scalar field in the infrared sector
can be written as a Fourier expansion in terms of the modes 
$\chi_k =e^{{\rm i} \vec k. \vec x} \xi_k(t)$
\begin{equation}
\chi_{cg}(\vec x,t) = \frac{1}{(2\pi)^{3/2}} {\Large \int} d^3k \ \theta(k-
\epsilon k_0) \left[a_k \chi_k + a^{\dagger}_k \chi^*_k \right],
\end{equation}
where $a_k$ and $a^{\dagger}_k$ are the anihilation and creation
operators, which satisfy  
the commutation relations\cite{BCMS}:
$\left[a_k, a^{\dagger}_{k'}\right] = \delta^{(3)}(k-k')$,
$\left[a^{\dagger}_k, a^{\dagger}_{k'}\right]=\left[
a_k, a_{k'}\right]=0$. Furthermore, $\epsilon \ll 1$ is a
dimensionless constant\cite{BCMS}, which is of the order $\epsilon \simeq
10^{-3} - 10^{-4}$.

From eq. (\ref{c}), one can write the equation of motion for the 
time dependent modes $\xi_k$ as a function of the homogenous scalar
field $\phi_c$
\begin{equation}\label{d}
\xi''_k(\phi_c) + \left[ \frac{12 \pi k^2}{(m M_p)^2 a^2} 
- 4 M^2 \phi^2_c + 6 M\right] \xi_k(\phi_c) =0,
\end{equation}
where the prime denotes the derivative with respect to $\phi_c$
and $M={3\pi \over M^2_p}$.
Since we are interested in studying the super Hubble ($k^2 \ll k^2_0$)
matter field fluctuations, the equation (\ref{d}) takes the form
\begin{equation}\label{e}
\xi''_0- \left[ 4 M^2 \phi^2_c - 6 M\right] \xi_0 =0.
\end{equation}
The solutions of this equation describe the infrared sector and thus
must be real\cite{CQG99} to obtain classicality conditions of matter
field fluctuations in this sector. Hence, the real solution for $\xi_k$
comes out to be
\begin{equation}
\xi_0(\phi_c) = |c_1| \  e^{-M\phi^2_c} \  \phi_c ,
\end{equation}
where $c_1$ is an arbitrary constant. The squared amplitude for the
infrared matter field fluctuations $\phi_{cg} = a^{-3/2} \chi_{cg}$
are $\left< \phi^2_{cg}\right>_{IR} 
= {a^{-3} \over 2\pi^2} \int^{\epsilon k_0}_{0}
dk \  k^2 \  \xi^2_0$, so that
\begin{equation}
\left< \phi^2_{cg}\right>_{IR} \simeq \frac{c^2_1 \epsilon^3}{6\pi^2}
\left[ 4 M^2 \phi^2_c - 6 M\right]^{3/2} \phi^2_c \  e^{-2M \phi^2_c}.
\end{equation}
For inflation to take place one requires 
that $4M^2 \phi^2_c - 6 M >0$, so that 
$\phi^2_c > {M^2_p \over 2 \pi}$. Furthermore, the 
energy density fluctuations 
${\delta\rho \over \rho} \simeq 
{V'_c \over V_c} \  \left<\phi^2_{cg}\right>^{1/2}$ are
\begin{equation}
\frac{\delta\rho}{\rho} \simeq \frac{c_1 \epsilon^{3/2}}{\sqrt{6}\pi}
\left[ 4 M^2 \phi^2_c - 6 M\right]^{3/4} e^{-M \phi^2_c},
\end{equation}
which agree with the COBE data, ${\delta\rho \over \rho} \simeq 10^{-5}$, for 
$c_1 \simeq 208.71$.
The power spectrum is given by ${\cal P}_{\phi_{cg}}=\left|\delta_k\right|^2$,
for $\left<\phi^2_{cg}\right>_{IR}
=\int^{\epsilon k_0}_{0} 
{dk \over k} {\cal P}_{\phi_{cg}}$, such that the
spectral density $\delta_k$ for this model is
\begin{equation}
\delta_k = \frac{c_1 \epsilon^{3/2}}{\sqrt{6} \pi}  
\left[4 M^2 \phi^2_c - 6 M\right] \  e^{-M\phi^2_c} \  k^{3/2}.
\end{equation}
The spectrum $n-1=-6\gamma +2 \eta$ is given by [see eqs. (\ref{a})
and (\ref{b})]
\begin{equation}
|n-1| = \frac{3M^2_p}{2\pi} \  \phi^{-2}_c,
\end{equation}
which is very close to the Harrison - Zeldovich spectrum 
(i.e., $n=1$)\cite{HZ} for $\phi_c \gg M_p$
and is in good agreement with CMB data spectrum
$|n-1.2| < 0.3$\cite{PR}.

\subsection{Super Hubble
metric fluctuations for a massive scalar field}

Since $\phi_c(t) = \phi_i - {m M_p\over 2 \sqrt{3\pi}} t$, the
eq. (\ref{C}) for $\tilde\xi_0(\phi_c)$ can be written as
\begin{equation}
\tilde\xi''_0 - \left(4 L^2 \phi^2_c + 6 L\right) \tilde\xi_0 =0,
\end{equation}
where $L = {\pi \over M^2_p}$. This has the real solution
\begin{equation}
\tilde\xi_0(\phi_c) = 
e^{L\phi^2_c} \left[D_1\phi_c
+D_2 e^{-2L\phi^2_c}+ D_2 \  \sqrt{2\pi L} \  \phi_c \  {\rm Erf}
[\sqrt{2L} \  \phi_c]\right],
\end{equation}
where ($D_1$,$D_2$) are real constants and ${\rm Erf}$ is the error function.
Hence, the amplitude of fluctuations for the metric perturbations for 
super Hubble scales is
\begin{equation}
\left< \psi^2_{cg}\right> \simeq \frac{a^{-1}}{2\pi^2} 
{\Large\int}^{\epsilon k_0}_{0} dk \  k^2 \left[\tilde\xi_0(\phi_c)\right]^2,
\end{equation}
and the fluctuations for energy density are [see eq. (\ref{22})]
\begin{eqnarray}
\left.\frac{\delta\rho}{\rho}\right|_{IR} 
&\simeq & 2 \left<\psi^2_{cg}\right>^{1/2} \simeq
\frac{2 a \epsilon^{3/2}}{\sqrt{6} \pi}  \tilde\xi_0(\phi_c) \mu^{3/2} \nonumber
\\
&=& \frac{2 a_i e^{-L \phi^2_c}}{\sqrt{6} \pi} \epsilon^{3/2}
\left(4 L^2\phi^2_c + 6 L\right)^{3/4} \left[
D_1 \phi_c+ D_2 \left(e^{-2L\phi^2_c}+\sqrt{2\pi L} \  \phi_c \  {\rm Erf}[
\sqrt{2L} \  \phi_c]\right)\right],
\end{eqnarray}
which increase when $\phi_c$ decreases (i.e., when the time increases).
This means that super Hubble scalar metric fluctuations increase
during inflation for a massive scalar field.

\section{Final Comments}

In this work we studied the evolution of both, super Hubble matter and
metric fluctuations during inflation. We have considered the case
of a massive scalar field with a potential given by  
$V(\phi_c) = {m^2 \over 2} \phi^2_c$.
The fact that the modes for the redefined fields $\chi_{cg}$ and $h_{cg}$
can be mapped
as functions of $\phi_c$, makes it possible to find solutions
for the zero modes ($k=0$) for $\xi_k$ and $\tilde\xi_k$, and thus 
$\left<\phi^2_{cg}\right>$ and $\left<\psi^2_{cg}\right>$ can be
calculated by means of a semiclassical method.

We find that super Hubble
matter field fluctuations increase during inflation so that
at the end of inflation 
$\delta\rho/\rho \simeq 10^{-5}$ is reached. 
Furthermore, we get 
a spectral index $n\simeq 1$, which describes a nearly scale invariant
spectrum for super Hubble matter field fluctuations.
Finally, we find that
scalar metric fluctuations increase during inflation. This behaviour 
agrees with  results from a previous work in which 
the IR scalar metric perturbations are studied for a power-law
expanding universe\cite{pr2000}. It is interesting to point out, 
however, that metric perturbations are not amplified after 
$m^2  \phi^2$ inflation. As it has been recently shown\cite{reheating}, 
they cease to grow during reheating. 

\section*{Acknowledgements}

F. Astorga and M. Bellini would like to thank Conacyt  
and CIC of Universidad Michoacana for financial
support in the form of a research grant.

\end{document}